\documentclass[sigplan,10pt]{acmart}

\usepackage[ruled,vlined,linesnumbered]{algorithm2e} 
\usepackage{listings}
\usepackage{graphicx}
\usepackage{lineno}
\usepackage[official]{eurosym}
\usepackage[utf8]{inputenc}
\usepackage{textcomp}
\usepackage{pgfplots}
\usepackage{xparse}
\usepackage[utf8]{inputenc}
\usepackage[english]{babel}
\usepackage{amsthm}
\usepackage{amsmath}
\usepackage[caption=false]{subfig}
\usepackage{hyperref}
\usepackage{wrapfig}
\usepackage[subtle]{savetrees}

\newtheorem{theorem}{Theorem}[section]
\newtheorem{lemma}[theorem]{Lemma}
\newtheorem{definition}{Definition}[section]

\usetikzlibrary{patterns}
\pgfplotsset{compat=1.14}
\usepgfplotslibrary{
  units,
  groupplots
}

\newcommand{\orderlesschainspace}{\textsc{OrderlessChain }}
\newcommand{\orderlesschain}{\textsc{OrderlessChain}}

\begin{document}

\title{\orderlesschain: Do Permissioned Blockchains Need Total Global Order of Transactions?}

\titlenote{\color{blue} This paper is a preprint of the work published at the 24th ACM International Middleware Conference (Middleware ’23). DOI: https://doi.org/10.1145/3590140.3629111. Please cite the original Middleware conference version of the paper. }

\author{Pezhman Nasirifard}
\affiliation{
  \institution{Technical University of Munich}
   \country{Germany}
}
\email{p.nasirifard@tum.de}

\author{Ruben Mayer}
\affiliation{
  \institution{Technical University of Munich}
   \country{Germany}
}

\author{Hans-Arno Jacobsen}
\affiliation{
  \institution{University of Toronto}
  \country{Canada}
}

\renewcommand{\shortauthors}{P. Nasirifard, et al.}

\begin{abstract}
Existing permissioned blockchains often rely on coordination-based consensus protocols to ensure the safe execution of applications in a Byzantine environment. Furthermore, these protocols serialize the transactions by ordering them into a total global order. The serializability preserves the correctness of the application's state stored on the blockchain. However, using coordination-based protocols to attain the global order of transactions can limit the throughput and induce high latency. In contrast, application-level correctness requirements exist that are not dependent on the order of transactions, known as \textit{invariant-confluence (I-confluence)}. The I-confluent applications can execute in a coordination-free manner benefiting from the improved performance compared to the coordination-based approaches. The safety and liveness of I-confluent applications are studied in non-Byzantine environments, but the correct execution of such applications remains a challenge in Byzantine coordination-free environments. This work introduces \orderlesschain, a coordination-free permissioned blockchain for the safe and live execution of I-confluent applications in a Byzantine environment. We implemented a prototype of our system, and our evaluation results demonstrate that our coordination-free approach performs better than coordination-based blockchains.
\end{abstract}


\settopmatter{printacmref=false}
\setcopyright{none}
\renewcommand\footnotetextcopyrightpermission[1]{}
\pagestyle{plain}
\settopmatter{printfolios=true}

\maketitle

\vspace{-0.2cm}

\section{Introduction} \label{introduction} 

The main property contributing to blockchains' popularity since the introduction of Bitcoin~\cite{bitcoin} is the trusted execution of transactions in a trustless, decentralized environment. To offer trust and prevent Byzantine behaviors such as Sybil attacks~\cite{sybil, sybil_2}, blockchains use consensus protocols, such as the Proof-of-Work-based (PoW) protocols used in Bitcoin~\cite{bitcoin}. Another essential property of consensus protocols enables the system to agree on the total global order of transactions for a serialized execution. The serializability is required to preserve the correctness of the application's state stored on the blockchain. For example, serialization prevents a user's negative account balance in the case of Bitcoin, as every node sequentially executes the transactions in the same order. However, the consensus protocols in several blockchains are severe bottlenecks to their throughput and latency~\cite{survey_bc, scale_bc}. 

In contrast to public blockchains, permissioned blockchains are only accessible by authenticated and authorized participants~\cite{survey_bc, bc_wild}. Although the participants' identity is known, they do not trust each other. Permissioned blockchains, such as \textit{Hyperledger Fabric (Fabric)}~\cite{mainfabric}, take advantage of their permissioned property to implement more efficient coordination-based consensus protocols. However, the coordination-based nature of these protocols remains a bottleneck~\cite{fastFabric, databasify, jeeta, jeeta_2}.

In general, decreasing coordination plays a vital role in improving the performance of distributed systems~\cite{confluen}. A coordination-free blockchain could enable the concurrent execution of transactions, leading to higher throughput and lower latency. However, simply eliminating the coordination may jeopardize the correctness depending on the application. For example, a payment processing application may require rejecting transactions that result in the payee's account's balance turning negative~\cite{bc_payment}. A coordination-free blockchain cannot preserve this requirement~\cite{confluen, martin_bft}. 

In contrast, there exist application-level correctness requirements that \emph{can} be preserved in a coordination-free distributed system, which are known as \textit{Invariant-Confluent (I-confluent)} invariant conditions~\cite{confluen}. For example, transactions that only deposit funds to an account can execute without coordination. In other words, the I-confluent transactions can be processed in any order while preserving application-level correctness, and the final state of the application is independent of the order of the transactions. One technique that can create I-confluent transactions is \textit{Conflict-free Replicated Data Types (CRDTs)}~\cite{first_crdt}. CRDTs are abstract data types that converge to the same state in a coordination-free environment. 

Bailis et al.~\cite{confluen} demonstrated that unordered transactions preserve the I-confluent invariants of applications in non-Byzantine and eventually consistent environments. In other words, applications with I-confluent invariants are safe and live in non-Byzantine coordination-free environments. The authors also showed the improved throughput and latency of taking advantage of coordination-free approaches. However, preserving the safety and liveness of applications in a Byzantine environment is dependent on paying a high coordination cost in other systems~\cite{fastFabric, bft_crdt_1, scale_bc, survey_bc, databasify, jeeta}. By providing a Byzantine coordination-free environment where I-confluent applications continue to be safe and live, we benefit from improved performance while ensuring trust in a trustless environment. In this work, we present \orderlesschain, a coordination-free permissioned blockchain without total global order of transactions. \orderlesschainspace uses the properties of permissioned blockchains and CRDTs to offer an innovative two-phase execute-commit protocol for creating safe and live applications in a Byzantine coordination-free environment. We also built five applications on \orderlesschainspace to show its practicability.

We offer the following contributions in this paper:

\begin{enumerate}

\item We introduce \orderlesschain, a novel permissioned blockchain capable of executing safe and live applications in a Byzantine coordination-free environment. Our system achieves this without the coordination overhead for creating a total global order of transactions, improving the throughput and scalability. 

\item We demonstrate a novel approach for creating Turing-complete blockchain applications based on CRDTs. Our approach preserves the I-confluent invariants of applications in a coordination-free Byzantine environment. 

\item We implement a prototype of \orderlesschainspace and demonstrate the improved throughput and latency of the coordination-free approach for I-confluent applications compared to existing permissioned blockchains. 

\end{enumerate}

The remainder of the paper is organized as follows. First, we provide a background on I-confluence and CRDTs in Section~\ref{background} followed by the system model in Section~\ref{preliminaries}. Then, we explain our protocol in Section~\ref{protocol}. We discuss the applications of \orderlesschainspace and the implementation in Sections~\ref{orderless_apps} and \ref{impel}. We also explain our approach for preserving application-level correctness requirements in Section~\ref{invariant_cond} and the effects of Byzantine participants in Section~\ref{byzantine_actors}. We present evaluations in Section~\ref{evaluation} and review related work in Section~\ref{rw_work}.

\section{Background} \label{background}

\textbf{Invariant Conditions and Invariant Confluence --} Different applications have different correctness requirements. For example, a banking application may be required to prevent the customers' account balances from dropping below zero. Developers specify the correctness of an application by defining a set of invariant conditions $\mathit{\{I_{1}, ..., I_{s}\}}$ on the application's state. Each $\mathit{I_{j}}$ represents a requirement that nodes must preserve during the application's lifecycle. Preserving invariants in a distributed system with globally serialized transactions is relatively straightforward. Provided that each transaction preserves the invariants, serialization enables the nodes to apply the transactions in a sequentially isolated manner and preserve the invariants. 

However, serialization comes at a high coordination cost. In a coordination-free distributed system, the nodes may receive the transactions in different orders. Hence, preserving invariants is challenging. For example, a node that stores the account balance of a customer with an account balance of $\mathit{\{Balance: 100\} }$ can accept only one of the withdrawal transactions of  $\mathit{Withdraw(50)}$ and $\mathit{Withdraw(60)}$. Applying both transactions results in a negative account balance and violates the application's invariants. Without coordination, the nodes cannot agree to accept one of the two transactions. 

Bailis et al.~\cite{confluen} studied preserving invariants in a non-Byzantine coordination-free distributed system and introduced the notion of \textit{Invariant Confluence (I-confluence)}. A set of transactions $\mathit{\{TS_{1}, ..., TS_{m}\}}$ are I-confluent with regard to an invariant condition $\mathit{I_{j}}$, if the transactions can be applied in different orders on different nodes while preserving $\mathit{I_{j}}$. Consider the mentioned withdrawal transactions as an example of a non-I-confluent transaction set. However, two deposit transactions $\mathit{Deposit(50)}$ and $\mathit{Deposit(60)}$ are I-confluent, as applying these transactions in any order on different nodes does not violate the non-negative invariant condition. Hence, the I-confluent transactions must have these two properties: (1) \textit{Commutativity:} The transactions can be applied in any order. (2) \textit{Convergence:} The final state is independent of the order of transactions. Bailis et al. proved that only I-confluent transactions could be executed on a coordination-free distributed system and non-I-confluent transactions require coordination among the system's nodes~\cite{confluen}.

\textbf{Conflict-free Replicated Data Types --}  One available technique that provides commutative and convergent transactions as required by I-confluence is Conflict-free Replicated Data Types (CRDTs). CRDTs represent abstract data types that converge to the same state in the presence of concurrent transactions in a coordination-free distributed system~\cite{first_crdt}. These data types encapsulate common data structures such as maps and provide APIs for reading and modifying their values. Since concurrent transactions can result in conflicting values, CRDTs use built-in mechanisms to resolve conflicts without coordination. Shapiro et al.~\cite{first_crdt} formalized CRDTs and proved their strong eventual consistency property (SEC) in an eventually consistent system. An SEC system has two requirements: (1) \textit{Eventual delivery of transactions:} If a transaction is delivered to one correct node, then all correct nodes will eventually receive the transaction. (2) \textit{Strong convergence of nodes:} If the same set of transactions is applied on every correct node, then the nodes' state immediately converges to the same state~\cite{first_crdt}. 

CRDTs synchronize among different nodes through propagating commutative transactions~\cite{Fast_as_Possible}. When extending common data structures with CRDT features, the transactions may inherently be commutative or not. For example, a counter is easily modeled as a CRDT since increment transactions are intrinsically commutative. However, modifications for several other data types are not commutative. For instance, assigning a value to a single-value register is not inherently commutative. For converting a register to a CRDT, the register needs to be extended with metadata, defining its behavior in the presence of concurrent modifications. This is achieved with the help of the \textit{happened-before} relation~\cite{first_crdt} that defines the causal order between two events based on \textit{logical clocks}~\cite{lamport}. The theoretical foundation for defining the requirements of several CRDTs has been studied thoroughly~\cite{crdt_overview, jsoncrdt}.

\section{System Model} \label{preliminaries}

\textbf{System Model --}  \orderlesschainspace is a strongly eventually consistent, asynchronous permissioned blockchain. An \orderlesschainspace network consists of a set of organizations $\mathit{\{O_{1}, ..., O_{n}\}}$ and a set of clients $\mathit{\{C_{1}, ..., C_{r}\}}$. Organizations can communicate with other non-failed organizations by sending and receiving messages. 

A unique identifier is assigned to each organization and client. The identity of each organization is known to every other organization and client in the network. An organization represents entities that range from large corporations to small businesses or even individuals. The purpose of organizations is to define trust boundaries in the system. Although the organizations' identity is known to each other, the organizations do not necessarily trust each other.  

\textbf{Running Example --}  To better convey our system model and design, we create a voting application, to which we refer throughout the paper. Each voter $\mathit{Voter_i}$ can vote for one party among the candidate parties in $\mathit{\{P_{1}, ..., P_{n}\}}$. The network consists of $\mathit{n}$ organizations, where each organization represents one distinct party. Each organization receives and stores votes from voters. We consider the application correct if each voter votes for at most one party. We chose this use case since voting applications are among popular blockchain use cases~\cite{voting}. Additionally, studies have shown that coordination in such highly concurrent use cases is a bottleneck~\cite{jeeta, databasify}.  

\textbf{Application's World State --} Each organization stores a replica of the application's state as a set of key-value pairs represented by $\mathit{ST_{O_{i}}}$, which represents the application state at organization  $\mathit{O_{i}}$. Since \orderlesschainspace is an SEC system, the replicated application states $\mathit{ST_{O_{1}}, ..., ST_{O_{n}} }$ at organizations $\mathit{O_{1}, ..., O_{n}}$ may diverge from each other, but will eventually converge to the same state. At any given point in time, we define the application's world state $\mathit{ST_{App}}$  as $\mathit{ST_{App} = \cup_{i=1}^{n}ST_{O_{i}}}$  as the union of the application state at all organizations where the values of identical keys are merged based on the techniques discussed in this paper.

\textbf{Invariant Conditions --} An application's correctness is imposed by the developer by defining a set of invariant conditions $\mathit{\{I_{1}, ..., I_{s}\}}$ on  $\mathit{ST_{App}}$. Each invariant $\mathit{I_j}$ specifies a constraint over $\mathit{ST_{App}}$. We define the application correctness as follows:

\begin{definition}{\textbf{$\mathbf{ST_{App}}$ Correctness.}}
Let $\mathit{ST_{App}}$ be the application's world state that does not violate the invariant conditions $\mathit{\{I_{1}, ..., I_{s}\}}$. Let the transaction set $\mathit{\{TS_{1}, ..., TS_{m}\}}$ be I-confluent with regard to $\mathit{\{I_{1}, ..., I_{s}\}}$. Then, committing the transactions $\mathit{\{TS_{1}, ..., TS_{m}\}}$ does not violate any invariant conditions $\mathit{\{I_{1}, ..., I_{s}\}}$ over $\mathit{ST_{App}}$.
\end{definition}

\textbf{Application's Endorsement Policy --} The application developers specify the \textit{endorsement policy} for the application. The endorsement policy specifies which organizations must sign and commit the transactions. The process of obtaining the signature is called \textit{endorsing}. The application's endorsement policy has the format $\mathit{EP: \{ q \; of \; n\} }$, where $\mathit{n}$ is the number of organizations in the system, and $\mathit{q}$ is the minimum number of organizations required for endorsing as well as committing a transaction. In other words, the endorsement policy determines the trust requirements of the application and enables the developer to adjust the amount of trust required. \\

In the context of our voting example, consider an election with four participating parties $\mathit{P_1, P_2, P_3, P_4}$ where each party is represented by a corresponding organization $\mathit{O_{P_1}, O_{P_2}, O_{P_3}, O_{P_4}}$. Consider the following two possible endorsement policies:  $\mathit{EP_1: \{2 \; of \; 4 \} }$ and  $\mathit{EP_2: \{4 \; of \; 4 \} }$. $\mathit{EP_1}$ requires that votes are endorsed and committed by at least two out of the four organizations. $\mathit{EP_2}$ indicates that all four organizations must endorse and commit the voter's vote. Furthermore,  we identify one invariant condition: \textit{maximally one vote per voter}. The application is correct if the \textit{maximally one vote per voter} invariant is preserved over $\mathit{ST_{App}}$ and committing transactions do not violate this invariant. 

\textbf{Transaction Model --} A transaction is valid as follows:

\begin{definition}{\textbf{Transaction Validity.}}
Let the application's endorsement policy be $\mathit{EP: \{ q \; of \; n\} }$. Let  $\mathit{ST_{App}}$ be in a correct state concerning the invariant conditions. Let the transaction $\mathit{TS_{i}}$ be I-confluent concerning the invariant conditions. Then, $\mathit{TS_{i}}$ is \textit{valid} if and only if it satisfies these two requirements: (1) \textit{Signature validity:} $\mathit{TS_{i}}$ is endorsed by at least $\mathit{q}$ organizations and the client signed the transaction. (2) \textit{Invariant conditions validity:} Applying $\mathit{TS_{i}}$ does not violate any invariants. 
\end{definition}

We define the transaction $\mathit{TS_{i}}$ to be committed as follows:

\begin{definition}{\textbf{Committed Transaction.}}
Let the application's endorsement policy be $\mathit{EP: \{ q \; of \; n\} }$. Let the transaction $\mathit{TS_{i}}$ be valid. Then, $\mathit{TS_{i}}$ is successfully committed if and only if at least $\mathit{q}$ organizations individually process and commit the transaction successfully.
\end{definition} 

For the voting example with $\mathit{EP_1: \{2 \; of \; 4 \}}$, a transaction is valid if it is signed by the client and is endorsed by at least two organizations. Additionally, the valid transaction must not violate the \textit{maximally one vote per voter} invariant. Also, at least two organizations must commit a valid transaction. 

\textbf{Failure Model --} We consider the organizations and clients to be potentially Byzantine. Byzantine organizations or clients can fail arbitrarily. We consider an organization to be non-faulty if and only if the organization processes every transaction according to the \orderlesschain's protocol. The transactions can be delivered in any order, differing from the sent order; they may also be duplicated, lost, or corrupted during transmission. The safety and liveness properties of applications running on \orderlesschainspace are defined as follows:

\begin{definition}{\textbf{Safety.}}
Only valid transactions are successfully committed. 
\end{definition}

\begin{definition}{\textbf{Liveness.}}
Every valid transaction is eventually successfully committed. 
\end{definition}

We have two kinds of failures: (1) \textit{Signature failure:} When a transaction does not receive the required endorsements based on the endorsement policy, or the client's signature is not valid. (2) \textit{Organization failure:} Any Byzantine failures of the organizations, including crash and omission failures and the organizations' arbitrary behavior, such as intentionally jeopardizing the system through tempering with messages, forging signatures, or software bugs. 

Intuitively speaking, consider the two possible endorsement policies for our voting example. $\mathit{EP_1}$ requires the endorsement and committing of at least two organizations. Therefore, at most, one of four organizations can be Byzantine, so the other non-faulty organizations can prevent committing invalid transactions and keep the application safe. With more than one Byzantine organization, the client may collude with the Byzantine organizations and collect the two required endorsements and commits for the invalid transactions, and the non-faulty organizations cannot prevent it. However, the voting application with $\mathit{EP_2}$ is safe for up to three Byzantine organizations, as the remaining one non-faulty organization can prevent the successful commit of invalid transactions. For liveness with $\mathit{EP_1}$, the client must communicate with at least two organizations. As there are four organizations, liveness can tolerate two Byzantine failures. However, the liveness of $\mathit{EP_2}$ cannot tolerate any Byzantine failures, as any faulty organization can hinder the transaction from being endorsed or committed by all four organizations.

Formally speaking, for an application with the endorsement policy $\mathit{EP: \{ q \; of \; n\} }$ and with up to $\mathit{f}$ Byzantine organizations, the application is safe if $\mathit{q \geq f + 1}$. Additionally, the application is live, if $\mathit{n - q  \geq f }$. We provide proof of the safety and liveness of \orderlesschainspace in Section~\ref{byzantine_actors}. The safety and liveness condition of \orderlesschainspace in a Byzantine environment differs from the conventional $\mathit{ 3f + 1}$ requirement, as we do not require the organizations to coordinate to reach a consensus. Rather, we use the permissioned property of the system and the organizations' known identity to endorse the transactions, where consequently, the non-faulty organizations prevent endorsing and committing invalid transactions.

In the case of a network partition, an application with the endorsement policy of $\mathit{EP: \{ q \; of \; n\}}$ can remain available if the number of organizations in every partition satisfies the safety and liveness requirements. Hence, \orderlesschainspace is available under network partitions according to the CAP theorem~\cite{cap}, if in every partition there exists at least $\mathit{q}$ organizations, and once the network partition is resolved, the state of partitions can be merged based on the techniques discussed in this paper.

\section{Architecture and Protocol} \label{protocol}

\textbf{\orderlesschainspace Architecture --} Organizations are responsible for hosting smart contracts, receiving and executing transactions, and managing a replica of the application's ledger. Every application running on \orderlesschainspace makes use of an isolated ledger, which contains the application state $\mathit{ST_{O_{i}}}$. The application's ledger on every organization consists of two components: (1) an append-only hash-chain log; (2) a database. The hash-chain log contains all transactions that the organization has received since the beginning of time in a hash-chain data structure. By sequentially executing every transaction in the hash-chain log, we reach the application state $\mathit{ST_{O_{i}}}$. For a more efficient approach, an organization applies each transaction to its database when the transaction is appended to the log. Therefore, the database represents the current application state $\mathit{ST_{O_{i}}}$. 

The messages are authenticated using digital signatures based on a standard Public Key Infrastructure (PKI)~\cite{pki}. Organizations and clients use PKI to authenticate and sign transactions and verify the integrity of the messages.

Developers create smart contracts, which are programs containing the application's logic. \orderlesschainspace supports executing smart contracts with a Turing-complete logic written in the Go language~\cite{golang}. Each smart contract can contain any number of functions. Each function encapsulates a task that the application performs. To execute a smart contract, clients submit a request to an organization that executes the smart contract with the provided inputs and returns a result.

\begin{figure}[h!]
  \centering
  \includegraphics[width=1\linewidth]{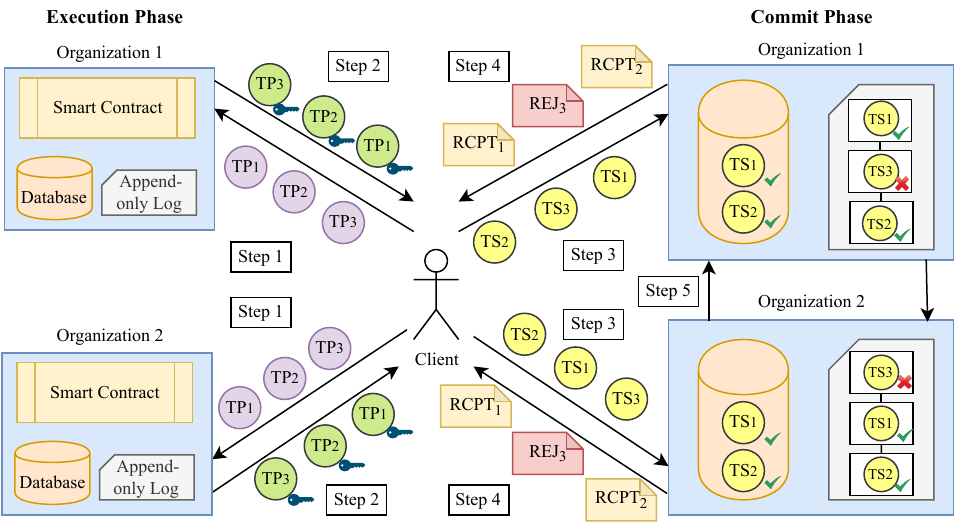}
  \caption{Transaction lifecycle on \orderlesschain.}
  \label{fig:orderlesschain}
\end{figure}

\textbf{Protocol and Transaction Lifecycle --} \orderlesschainspace follows a two-phase \textit{execute-commit} protocol and transaction lifecycle. Clients first submit transaction proposals to be executed by organizations. If the first phase succeeds, clients send the transactions to the organizations to be committed. Figure~\ref{fig:orderlesschain} demonstrates the complete transaction lifecycle for an application with endorsement policy $\mathit{EP: \{ q \; of \; n\} }$.

\textit{Phase 1 / Execution Phase --} The client prepares a transaction proposal $\mathit{TP_{i}}$ containing the client's identification, the smart contract's identifier, the function to be invoked, and the input parameters. The client broadcasts the proposal to at least $\mathit{q}$ organizations according to the endorsement policy (Step 1 in Figure~\ref{fig:orderlesschain}). Organizations receive the proposal and execute the smart contract with the provided parameters. The execution result is a set of I-confluent operations for modifying the application's state, created based on the CRDT methodology. These I-confluent operations preserve the application's invariant conditions, which we explain in detail in the following sections. The operations are added to a write-set. Then, the organization hashes and signs the write-set with its private key and creates a signature. Finally, the organization delivers the write-set with the created signature as a response (endorsement) to the client (Step 2 in Figure~\ref{fig:orderlesschain}). This signature ensures that the client or other organizations cannot tamper with the operations in the endorsement's write-set, as tampering makes the signature invalid.

\textit{Phase 2 / Commit Phase --} The client waits until it receives the minimum number of endorsements required by the endorsement policy. If the write-sets of all endorsements contain identical operations, the client assembles a transaction $\mathit{TS_{i}}$. The identical operations in the endorsements show that organizations followed the same protocol for executing the smart contract. Suppose some Byzantine organizations do not execute the smart contract defined by the developer or based on the provided input parameters. In that case, the operations will not match the operations created by non-Byzantine organizations and will cause the transaction to fail. The client adds the endorsement's write-set to the $\mathit{TS_{i}}$'s write-set. The client hashes and signs the transaction's write-set with its private key to create a signature and includes it in the transaction. The client also includes the received endorsements in the transaction.

The client sends back the transactions to at least $\mathit{q}$ organizations as specified by the endorsement policy (Step 3). These organizations can potentially be different from those who initially endorsed the proposal. If an organization has not previously committed the transaction, it validates and commits each received transaction according to the definitions above. Before committing a transaction, organizations verify whether the transaction's endorsements and client's signature are valid (\textit{signature validation}) and whether endorsements satisfy the endorsement policy. For verifying the validity of endorsements and the client's signature, the organization hashes the transaction's write-set and uses the public keys of endorsing organizations and the client to verify their signatures. This verification shows that the endorsing organizations created identical write-sets, and the client did not tamper with the write-set. If the transaction passes the signature validation, it is marked as valid. Otherwise, the transaction is invalid. 

The organizations update their database with the write-set of valid transactions, whereas all valid and invalid transactions are appended to the hash-chain log. For appending the transaction to the log, the organization creates a block $\mathit{Block_{h}: <TS_{i}, Hash(Block_{h-1})>}$, which contains the transaction and the hash of the last block $\mathit{Block_{h-1}}$ in the log. Then, the organization appends the created block to the log. For valid transactions, a receipt $\mathit{RCPT_{i}:}$ $\mathit{HashAndSign(Block_{h}, Valid)}$, that is signed hash of the block containing the transaction, is sent to the client (Step 4). If the transaction is invalid, the organization sends a rejection $\mathit{REJ_{i}: HashAndSign(Block_{h}, Invalid)}$ to the client. As the receipt contains the hash of the block, which is dependent on the hash of previous blocks in the log, the organization cannot modify the content of the transaction without destroying and invalidating $\mathit{RCPT_{i}}$ of $\mathit{TS_{i}}$ and other transactions. The client waits until it receives the minimum number of receipts required by the endorsement policy. The client can archive the transaction's receipts for bookkeeping purposes.

After sending the client's receipt, the organization periodically gossips the transactions to other organizations (Step 5). Upon receiving a transaction from another organization, the organization checks the ledger to determine if the transaction has already been received from other organizations or the client. If the transaction has already been processed, the organization ignores it and avoids committing it again; otherwise, it is committed following the above-explained procedure. If a client sends a transaction that the organization has received from other organizations or a duplicate transaction from the client itself, it does not commit it again. Instead, a receipt or rejection is sent to the client.

\section{\orderlesschainspace Applications} \label{orderless_apps}

By discussing two use cases, we explain the possible use cases of \orderlesschainspace and the system's internal approach for creating CRDT-based  I-confluent applications. 

\begin{figure}[h!]
  \centering
  \includegraphics[width=1\linewidth]{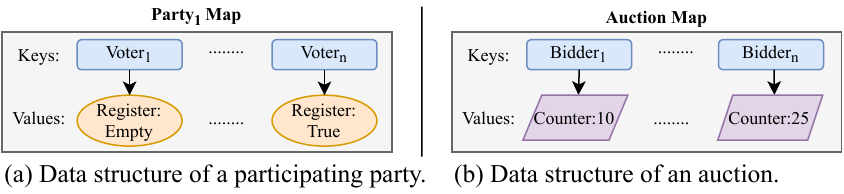}
\caption{Application modeling for the voting and auction.}
\label{fig:maps}
\end{figure}

\textbf{Application Modeling --} To implement a use case in a smart contract, we need to model the application as data structures that match the use case's description and contain the application's data. We discuss modeling two use cases:

\textit{Voting Application --} One possible solution for modeling our running voting example in a smart contract is shown in Figure~\ref{fig:maps}(a): For every party participating in the election, we require a map containing key-value pairs. The key is the voter's identification, and the value is a register that stores a Boolean value for the vote sent by the voter for this party. 

\textit{Auction Application --} Auction applications are among the common use cases of blockchains~\cite{auction}. An auction is a highly concurrent use case that can benefit from a coordination-free approach. Consider an auction and a set of bidders $\mathit{\{Bidder_1,}$  $\mathit{..., Bidder_n\}}$. The bidder $\mathit{Bidder_i}$ submits bids. Each bid contains the amount it wishes to add to its previous bid. The bidder must be able only to increase its last bid. Based on this description, we realize one invariant condition: \textit{increase-only bids}.

One possible design is as follows, as shown in Figure~\ref{fig:maps}(b): Each auction is modeled as a map containing key-value pairs. The key is the bidder's identification, and the value is a counter. The counter stores the cumulative bids of the bidders. The counter's value can only be increased, and the value is increased with every new bid sent by the bidder.

\textbf{CRDT Abstractions --} As explained in Section~\ref{background}, CRDTs provide a solution for creating commutative convergent update operations, and we use CRDTs in smart contracts. The proposed \orderlesschainspace protocol is independent of CRDTs used in smart contracts. CRDTs are also replaceable with alternative techniques that provide commutable operations to develop new types of applications. However, there exists a plethora of CRDTs for various data types. To enable the execution of these CRDTs, their specifications need to be supported by the smart contract execution environment. In the current implementation, \orderlesschainspace supports the specifications of grow-only counters (G-Counter)~\cite{first_crdt}, CRDT Maps~\cite{jsoncrdt}, and multi-value registers (MV-Register)~\cite{jsoncrdt}. We chose these three CRDTs as they satisfy the requirements of the voting and auction applications. Other use cases may require further CRDTs. For enabling the support for other CRDTs, their design requirements, according to the available literature, must be added to \orderlesschain~\cite{crdt_overview, first_crdt, crdt_text_1}.

\begin{table}[h!]
	\footnotesize
	\centering
	\caption{Modification and read APIs of supported CRDTs.}
	\begin{tabular}{| l l l |} 
		\hline
		\textbf{CRDT} & \textbf{Modification APIs}               & \textbf{Read API }     \\ 
		\hline
		\hline
		G-Counter     & $\mathit{AddValue(value, clock)}$        & $\mathit{Read()}$      \\ 
		\hline
		CRDT Map      & $\mathit{InsertValue(key, value, clock)}$ & $\mathit{Read(key)}$   \\
		\hline
		MV-Register   & $\mathit{AssignValue(value, clock)}$     & $\mathit{Read()}$ \\
		\hline
	\end{tabular}
	\label{table:crdt_apis}
\end{table}

The three CRDTs represent the following data structures: (1) \textit{G-Counter:} It is a monotonically increasing numeric variable. (2) \textit{CRDT Map:} This CRDT is built upon a map data structure. A map is an unordered data structure containing key-value pairs. The key is an identifier, and the value can be any object. (3) \textit{MV-Register:} This is a shared variable capable of containing multiple values at a time. Every CRDT provides modification and read APIs as shown in Table~\ref{table:crdt_apis}. Using the read APIs in the smart contracts causes no side effects and requires no CRDT operation. The developers create operations in the smart contract containing the modification API calls. The value must be \texttt{null} for deleting a value with modification APIs of CRDT Map and MV-Register. The modification APIs contain a logical clock used to infer the happened-before relations. For creating more complex data structures, maps can be nested, where the value of the key-value pairs can be either a new CRDT Map, G-Counter, or MV-Register.

These CRDTs are used for voting and auction applications as follows. \textit{Voting application:} As previously shown in Figure~\ref{fig:maps}(a), each party is modeled as a map, and the voter's votes are modeled as key-value pairs in the party's map where the values are registers. Therefore, we use a CRDT Map to model the party's map and the MV-Register as the votes' register. \textit{Auction application:} In the modeled auction application shown in Figure~\ref{fig:maps}(b), we use a map for modeling the auction and increase-only counters for bids. Hence, we use a CRDT Map to model the auction's map and G-Counters to model the bids of each bidder.

\begin{figure}[h!]
  \centering
  \includegraphics[width=1\linewidth]{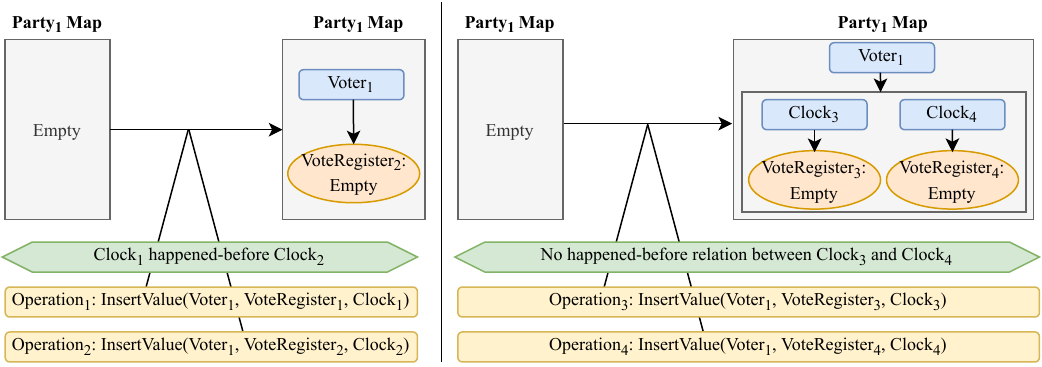}
\caption{Applying CRDT Map modification operations.}
\label{fig:merge_map}
\end{figure}

\begin{figure}[h!]
  \centering
  \includegraphics[width=1\linewidth]{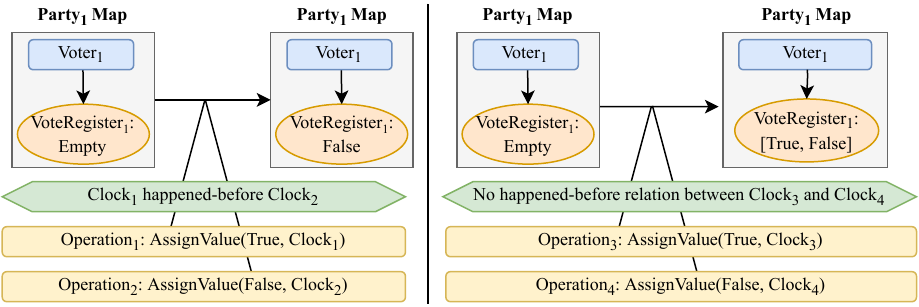}
\caption{Applying MV-Register modification operations.}
\label{fig:merge_register}
\end{figure}

For evaluating the effects of an operation, the operation needs to be applied to the CRDT, which may cause conflicts. The CRDTs must provide a built-in mechanism for resolving conflicts of modification operations. We identify the conflicting operations of the three CRDTs and offer a conflict resolution accordingly. (1) \textit{G-Counter:} As the operations increase the counter's value, the modification operations are inherently commutative and cause no conflict. (2) \textit{CRDT Map:} The modification operations that modify different keys in the map are commutative and non-conflicting and can be applied concurrently. However, the operations that modify identical keys are conflicting. The conflict is resolved based on the happened-before relations among operations. If the happened-before relation can be inferred, the operations are applied based on the relation; however, if the happened-before relation cannot be inferred, a new map is created, and the conflicting values are added to the new map as new key-value pairs, as shown in Figure~\ref{fig:merge_map}. (3) \textit{MV-Register:} On MV-Register, every modification operation is conflicting, and the value of the register is determined based on the happened-before relation among clocks. However, if the happened-before relation cannot be inferred from the clocks, the register stores all values, as shown in Figure~\ref{fig:merge_register}.

\section{Implementation} \label{impel}

We implemented a prototype of \orderlesschainspace with the Go language~\cite{golang} and gRPC~\cite{grpc}. We open-sourced the code and the smart contracts discussed in this paper~\footnote{\url{https://github.com/orderlesschain/orderlesschain}} . 
 
\begin{figure}[h!]
    \centering
  \includegraphics[width=1\linewidth]{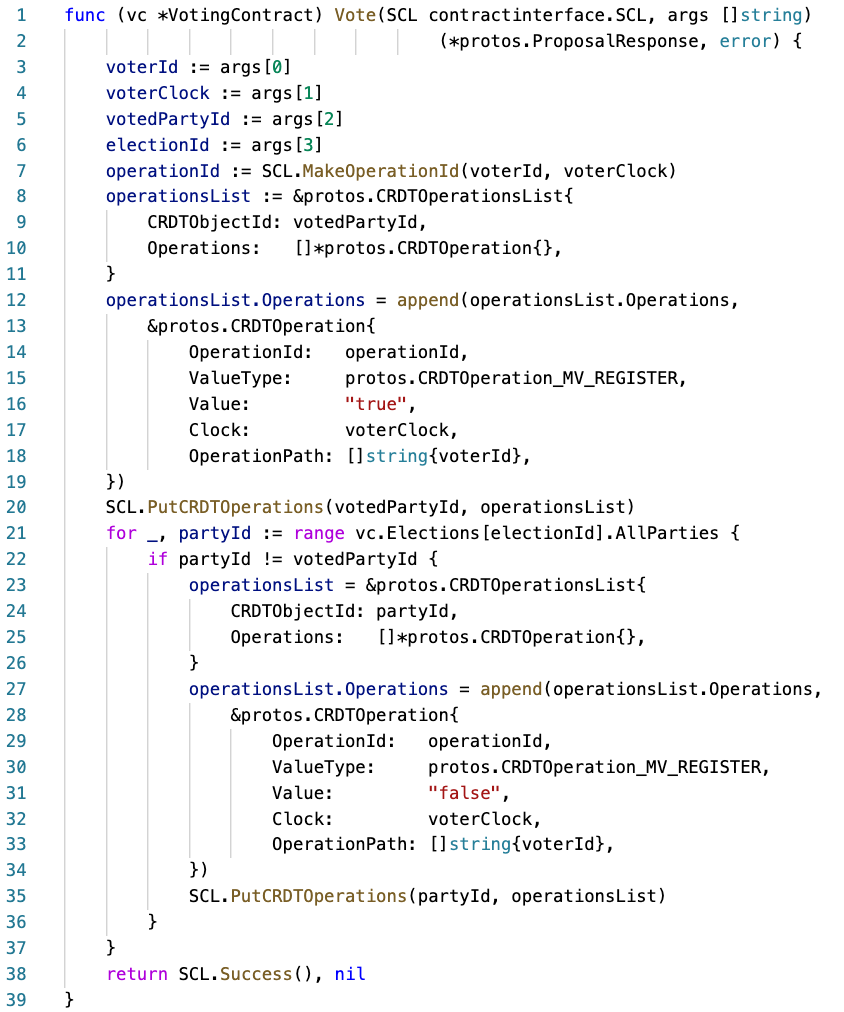}
  \caption{Voting in the smart contract.}
  \label{fig:code}
\end{figure}

\textbf{Smart Contracts --} Developers use our \textit{Smart Contract Library (SCL)} for developing smart contracts and defining the logic of applications. The smart contract includes functions that encapsulate different functionalities of the application. To enable developers to interact with data stored on the ledger, SCL offers interfaces for defining operations called CRDT APIs. Each client keeps track of a Lamport clock~\cite{lamport}, which is passed into the smart contract with proposals. The client increments the clock with every submitted proposal. Each client's Lamport clock is independent of the clock of other clients. Furthermore, each CRDT object has a unique identification on the ledger. The read API does not require creating any operation, and SCL only requires the identification of the CRDT object to retrieve it. For modifications, in addition to the identification of the CRDT object, each operation includes four components: (1) \textit{Operation identifier:} The identification of the operation is unique per CRDT object and is a combination of the client's identification and the client's Lamport clock. (2) \textit{Modification value and type:} The value that the operation modifies and the type of CRDT. (3) \textit{Client's clock:} The client's Lamport clock. (4) \textit{Operation path:} Developers can create nested CRDT structures for creating more complex data structures. The path specifies the location of the modification, starting from the root of the CRDT object. In the voting application with four parties, Figure~\ref{fig:code} shows the function in the smart contract for creating the operations for voting for a party. For example, the function creates four operations for voting for party $\mathit{P_1}$. One operation sets the voter's MV-Register on party $\mathit{P_1}$ to \textit{true}, and the other three operations set the voter's MV-Register on the other three parties to \textit{false}. These four operations are included in the write-set of proposals for submitting a vote.

\textbf{Applying Transactions --} Developers can implement functions in smart contracts for invoking read APIs and retrieving the values of CRDT objects. Subsequently, clients can submit proposals to an organization $\mathit{O_i}$ for reading the values. In our voting example, the developer can implement a function to read the number of votes submitted to a party. As \orderlesschainspace is an SEC system, the application state  $\mathit{ST_{O_{i}}}$ may diverge from the application states on other organizations. Therefore, reading the values at $\mathit{O_i}$ only reflects the modifications applied at $\mathit{O_i}$. 

To compute the CRDT object's value in response to read API calls, the organization should retrieve and apply every operation in the ledger submitted for the CRDT object. As the number of operations increases, the time required for applying operations also increases. This increasing overhead is a well-known problem of CRDTs~\cite{crdt_problem_1, crdt_problem_2}. Hence, we implemented an optimization to address this issue. As Section~\ref{protocol} explains, the ledger contains a database beside the hash-chain log. The database is updated with every valid transaction. It consists of a conventional key-value database, namely \textit{LevelDB~\cite{levelDB}}, and an in-memory cache. Upon the transaction commit, the operations are inserted into LevelDB. We do so as retrieving the operations from LevelDB is more efficient than retrieving them from the log during a cache miss. The value of the CRDT object in the cache is updated with the transaction's operations according to Algorithm~\ref{alg:apply}. In response to read API calls, the organizations return the value of the CRDT object from the cache. This approach offers \textit{read-your-writes consistency} from the client's point of view~\cite{readyourwrite}.

Algorithm~\ref{alg:apply} demonstrates our approach for applying each operation to the CRDT object. For every operation, before applying it, the CRDT object is traversed from its root until it reaches the location defined by the operation's path (Line 3). As the object can be a nested structure, parts of the path might not have been added to the object yet. Therefore, the missing parts are created and added. Additionally, the location contains the clocks of the previously applied operations. Once the location for modification is reached (Line 4), the changes are applied (Line 5). For applying the changes, as we explained in the CRDT abstractions, the built-in conflict resolution is applied depending on the type of object and the clocks of previously applied operations. Additionally, the operation's clock is appended to the location's clocks. The time and space complexity of Algorithm~\ref{alg:apply} is $\mathit{O(n)}$, where $n$ is the number of operations being applied.

\begin{algorithm}[h!]
	\footnotesize
\SetKwInOut{Input}{input}\SetKwInOut{Output}{output}
 \caption{Applying operations to the CRDT.}
    \label{alg:apply}
 \SetKwProg{ApplyOperations}{ApplyOperations}{}{}
  \ApplyOperations{$\mathit{(CRDTObj, Operations)}$}{
    \Input{$\mathit{CRDTObj}$,  \textit{a reference to the CRDT object.} }
    \Input{$\mathit{Operations}$, \textit{the modification operations.} }
    
    \ForEach{$\mathit{Op_i}$ $\mathbf{in}$  $\mathit{Operations}$}{
    
     			$\mathit{CRDTObj.Create(Op_i.OpPath)}$     	
				 
				 $\mathit{Location = CRDTObj.GetModifyLoc(Op_i.OpPath)}$
				 
				  $\mathit{CRDTObj.Apply(Location, Op_i.Val, Op_i.ValType, Op_i.Clock)}$
    }
  }
\end{algorithm} 

In Section~\ref{byzantine_actors}, we prove the SEC property. However, first, we demonstrate that the application state $\mathit{ST_{O_{i}}}$ is independent of the order of transactions. We formulate the following lemma:

\begin{lemma}
\label{trans_final}
Independent of the processing order of transactions in the transaction set $\mathit{\{TS_{1}, ..., TS_{m}\}}$ in organization $\mathit{O_{i}}$, application state $\mathit{ST_{O_{i}}}$ converges to the same state for all  $\mathit{i}$.
\end{lemma}

\begin{proof}
The write-set of every transaction in $\mathit{\{TS_{1}, ..., TS_{m}\}}$ only contains CRDT modification operations. As CRDTs are provided with the built-in conflict resolution mechanism, applying the operations in the write-set of operations by using Algorithm~\ref{alg:apply} ensures that transactions can be processed in any order while converging to the same state. Hence, the convergence of $\mathit{ST_{O_{i}}}$ is independent of the order of transactions.
\end{proof}

\section{Preserving Invariant Conditions} \label{invariant_cond}

As explained in Section~\ref{background}, submitting a set of transactions $\mathit{\{TS_{1}, ..., TS_{m}\}}$ in a coordination-free approach preserves the invariants $\mathit{\{I_{1}, ..., I_{s}\}}$ if the set of transactions are I-confluent with regard to the invariants. Organizations can commit a set of I-confluent transactions without additional validations while preserving the invariants. Since the CRDT operations in the write-set of transactions modify the application's state, the operations must be I-confluent. As developers define the logic for creating operations in a smart contract, they must implement the identified invariants as I-confluent operations. 

In the case of our voting application, we realized the \textit{maximally one vote per voter} invariant. To determine that the invariant can be preserved by creating I-confluent operations, we reason as follows: Consider an election with two participating parties. As explained in Section~\ref{impel}, every transaction $\mathit{TS_{Vote}}$  that submits a vote has two operations in the write-set. One operation sets the voter's MV-Register in the elected party's map to \textit{true}. The other operation sets the voter's MV-Register for the non-elected parties to \textit{false}. 

\begin{figure}[h!]
  \centering
  \includegraphics[width=1\linewidth]{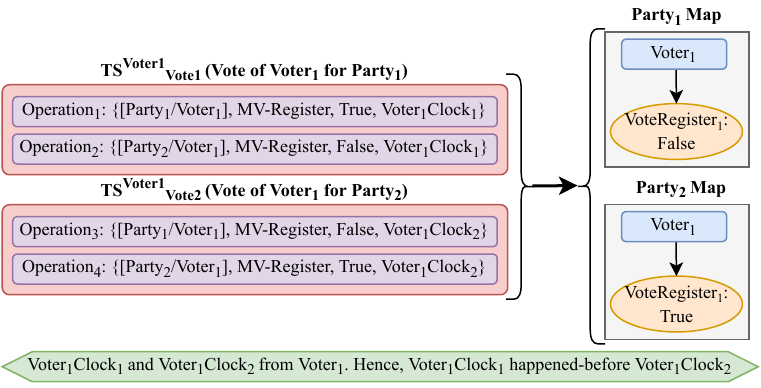}
\caption{Preserving the invariant for the voting application.}
\label{fig:iconf}
\end{figure}

As there is no coordination among organizations, the voter can submit several votes. However, the \textit{maximally one vote per voter} invariant requires that we only count one of the votes. Consider the following transaction set $\mathit{\{TS_{Vote1}^{Voter1}, TS_{Vote2}^{Voter1} \}}$, submitted by $\mathit{Voter_1}$, as shown in Figure~\ref{fig:iconf}. Each transaction contains two operations. $\mathit{Voter_1}$ submitted two votes for two different parties, where there exists a happened-before relation between operations in $\mathit{TS_{Vote1}^{Voter1}}$ and $\mathit{TS_{Vote2}^{Voter1}}$. Therefore, independent of the order they are processed, based on the CRDT's conflict resolution mechanism, operations in $\mathit{TS_{Vote2}^{Voter1}}$ overwrite the effects of operations in $\mathit{TS_{Vote1}^{Voter1}}$. Hence, we count only one of the votes submitted by the  $\mathit{Voter_1}$. The \textit{maximally one vote per voter} invariant is preserved, and the transactions are I-confluent with regard to the invariant. 

For the auction application, we can similarly reason that it is I-confluent concerning the \textit{increase-only bids} invariant. 

\section{Byzantine Actors} \label{byzantine_actors}

We assume that organizations or clients can be potentially Byzantine. We discuss four potential attacks by Byzantine clients: (1) A Byzantine client might send proposals to the organizations without submitting the transaction to be committed. This behavior does not leave any lasting side effects on the system. However, it can be used for distributed denial-of-service (DDoS) attacks. However, note that only authenticated clients can communicate with the organizations. Therefore, if the authenticated Byzantine clients try to overload the system, they can be detected, and their permissions can be revoked. (2) A Byzantine client may send the transactions to some organizations during the commit phase and avoid sending them to other organizations. As the organizations gossip the transactions to other organizations after committing the transaction, all organizations eventually receive the transactions. (3) Clients may send the wrong logical clocks to the organizations with the proposals. If clients send different clocks to different organizations with the same proposal, then the operations in the endorsements do not match, which prevents creating a valid transaction. (4) Suppose the client does not increment the clock with every proposal. In that case, no happened-before relation between clocks can be inferred, and the CRDT's conflict resolution mechanism manages the operations accordingly, as explained in Section~\ref{orderless_apps}. Therefore, Byzantine clients cannot jeopardize the system.

To discuss the safety and liveness concerning Byzantine organizations, we introduce the following theorem:

\begin{theorem} 
\label{safe_live}
Let the endorsement policy for an application be $\mathit{EP: \{ q \; of \; n \} }$ with $\mathit{ n \geq q > 0}$. Then, for up to $\mathit{f}$ Byzantine organizations, the application is safe if and only if $\mathit{q \geq f + 1}$. Furthermore, the application is live if and only if $\mathit{n - q  \geq f }$.
\end{theorem}

\begin{proof}
According to our definition of safety and liveness, the safe and live \orderlesschainspace must prevent committing invalid transactions and eventually commit valid transactions. 

Byzantine organizations may attempt to jeopardize the system by either responding with wrong messages or avoiding responding altogether. Wrong messages include forged signatures from organizations and clients, transactions with tampered or corrupted write-set operations, incorrectly executed smart contracts, or duplicated or lost messages. As the integrity of messages sent by organizations and clients can be examined, the signatures cannot be forged, and the organizations can independently prove the validity of organizations' and clients' signatures. As the system commits every transaction only once, and multiple executions of proposals do not leave any lasting side effects, duplication of messages has no effect. If the messages are suspected to be lost, they can be resent. Additionally, if a client's transaction fails due to the Byzantine organizations' wrong messages, the client can resubmit the proposals to another set of organizations and resend the transaction. On \orderlesschain, the developers identify and define the application logic for creating I-confluent update operations. Therefore, the invariants are preserved as long as the write-set operations are not tampered with and the smart contract is executed as defined by the developer. As the write-set of every endorsement must include identical operations, as long as there exists at least one non-faulty organization among the $\mathit{q}$ endorsing organizations, which creates the write-set operations that can be differentiated from the tampered operations or the incorrectly executed smart contact, creating a valid transaction is impossible, and the application is safe. Hence, the application is safe if and only if $\mathit{q \geq f + 1}$.

Byzantine organizations may not respond to clients. For the application to be live, the client must endorse and commit the transaction on $\mathit{q}$ among $\mathit{n}$ organizations. Therefore, the transaction can reach at least $\mathit{q}$ organizations if and only if $\mathit{n - q  \geq f }$. Therefore, the application is live if and only if $\mathit{n  - q  \geq f }$.
\end{proof}

We demonstrated that liveness and safety depend on the application's endorsement policy. In other words, the safety and liveness can be tailored to the application's requirements. For example, for the voting application with four parties, the regulation of a fair election may dictate that all parties endorse every vote. Therefore, we need $\mathit{EP: \{4 \; of \; 4\} }$. If the regulations demand the endorsement of at most two parties, we can have an $\mathit{EP: \{ 2 \; of \; 4\} }$. 

Furthermore, since the Byzantine behavior of organizations can be observed, and the identity of organizations is known to each other, the organizations have an incentive to behave honestly, as otherwise they may face the consequences. For example, a Byzantine party jeopardizing the election may face legal consequences.  

The following theorem demonstrates that $\mathit{ST_{App}}$ is SEC.

\begin{theorem}
Let the application be safe and live. Then, the application's world state $\mathit{ST_{App}}$ is SEC.
\end{theorem}

\begin{proof}
According to the definition of SEC in Section~\ref{background}, an SEC system must satisfy two requirements of \textit{eventual delivery of transactions} and \textit{strong convergence of nodes}. In Theorem~\ref{safe_live}, we demonstrated that every valid transaction is committed for a safe and live application. Additionally, non-faulty organizations gossip the transaction to other non-faulty organizations. Therefore, provided that the application is safe and live, every non-faulty organization eventually receives a valid transaction. Hence, \textit{eventual delivery of transactions} is satisfied. 

In Lemma~\ref{trans_final}, we proved that independent of the order of transactions in the transaction set $\mathit{\{TS_{1}, ..., TS_{m}\}}$, the application state $\mathit{ST_{O_{i}}}$ at organization $\mathit{O_{i}}$ converges to the same state for all $\mathit{i}$. As the \textit{eventual delivery of transactions} requirement for the safe and live application is satisfied, if the transaction set $\mathit{\{TS_{1}, ..., TS_{m}\}}$ is delivered to the non-faulty organization $\mathit{O_{i}}$, the same set is delivered to every other non-faulty organization. Therefore, according to Lemma~\ref{trans_final}, all $\mathit{ST_{O_{i}}}$ converges to the same state and the requirement \textit{strong convergence of nodes} is satisfied. Hence, the application's world state $\mathit{ST_{App}}$ of a safe and live application on \orderlesschainspace is SEC.
\end{proof}

\section{Evaluation} \label{evaluation}

We first evaluate \orderlesschain. Then, we compare it to \textit{Fabric}~\cite{mainfabric} and \textit{FabricCRDT}~\cite{FabricCRDT}. Fabric is one of the most prominent permissioned blockchains capable of executing Turing-complete applications similar to \orderlesschain. FabricCRDT (built as an extension on top of Fabric), to the best of our knowledge, is the only permissioned blockchain capable of running CRDT-enabled applications. Like \orderlesschain, Fabric's and FabricCRDT's network consists of organizations and uses endorsement policies. Unlike \orderlesschain, the clients send the transactions assembled from endorsements to an ordering service, creating a total global order of transactions by batching transactions into blocks and sending them to the organizations. Before the transaction commits, Fabric's organizations perform signature validation and a \textit{multi-version concurrency control validation (MVCC validation)} to ensure that the application's invariants are preserved. FabricCRDT only performs a signature validation and no MVCC validation and then merges the transaction values using JSON CRDT techniques~\cite{jsoncrdt}. 
 
We compare \orderlesschainspace to a prototype of Fabric and FabricCRDT, which we implemented based on the Go language, gRPC, and LevelDB. We do so because the original Fabric and FabricCRDT offer many security and network-related features that we do not provide in \orderlesschain. As these features impose performance penalties, we replaced the original implementations for the sake of fair comparison. The extensive evaluation of Fabric performed by Chacko et al.~\cite{jeeta} demonstrates these performance issues and confirms the fairness of our approach for using our prototypes of Fabric and FabricCRDT. Furthermore, the CRDT approach in FabricCRDT does not use the cache we implemented as an optimization. For fairness, we also implemented such a cache in FabricCRDT.

\textbf{Experimental Applications --} We developed a synthetic application for evaluating \orderlesschain. Based on the examples discussed, we also implemented voting and auction applications for comparing \orderlesschainspace to Fabric and FabricCRDT. Every application consists of one smart contract, and in total, we developed seven smart contracts. Each smart contract has one \textit{modify-function} for modifying the data on the ledger and one \textit{read-function} for retrieving data from the ledger. 

\textit{Synthetic Application --} For a controlled evaluation of \orderlesschain, we implemented a synthetic application. The application's smart contract includes two functions $\mathit{Modify(}$ $\mathit{ClientId_i, Clock_i, ObjCount, OpsPerObjCount, CRDTType)}$ and \\$\mathit{Read(ObjCount)}$. The $\mathit{Modify} $ function receives the client identification and clock, the number of CRDT objects to be modified, the number of operations per each CRDT object modification, and the CRDT type. CRDT type is either a G-Counter, CRDT Map, or MV-Register. The write-set of transaction includes $\mathit{ObjCount \times OpsPerObjCount}$ operations. The $\mathit{Read}$ function reads a specific number of CRDT objects as specified by $\mathit{ObjCount}$.

\textit{Voting Application --} We developed applications based on the voting example for \orderlesschain, Fabric, and FabricCRDT. The application's smart contract for \orderlesschainspace has two functions: $\mathit{Vote(Voter_i, Clock_i, Party_j, Election_l)}$ and $\mathit{ReadVoteCount(Party_j, Election_l)}$. For an election with $\mathit{n}$ parties, the $\mathit{Vote}$ function results in $\mathit{n}$ total operations (one operation per object) in the write-set as explained in Section~\ref{impel}. $\mathit{ReadVoteCount}$ retrieves the current number of votes of $\mathit{Party_j}$. The smart contracts for Fabric and FabricCRDT also include $\mathit{Vote}$ and $\mathit{ReadVoteCount}$ functions, which are implemented based on the best practices for developing smart contracts on these systems~\cite{jeeta,mainfabric, FabricCRDT}. 

\textit{Auction Application --} The auction applications are implemented for \orderlesschain, Fabric, and FabricCRDT. The application's smart contract of \orderlesschainspace has two functions: $\mathit{Bid(Bidder_i, Clock_i, }$ $\mathit{ BidIncrease_i,Auction_j)}$ and $\mathit{GetHighestBid(Auction_j)}$. The $\mathit{Bid}$ function includes one operation in its write-set for increasing the bidder's G-Counter. $\mathit{GetHighestBid}$ reads the current highest bid. The smart contracts for Fabric and FabricCRDT also includes a $\mathit{Bid}$ and a $\mathit{GetHighestBid}$ function.

\textbf{Workloads, Control Variables and Metrics --} Each experiment is executed on an initially empty ledger. We submit a workload containing transactions invoking the modify- and read-functions in the smart contracts, also referred to as \textit{modify-} and \textit{read-transactions}. The workload includes a specific percentage of modify-transactions and read-transactions, uniformly distributed during the execution of the experiment. Each organization receives a specific percentage of the load on the system. We define the transaction arrival rate as \textit{transactions per second (tps)} of the system as the total number of transactions per second submitted by all clients to the system. The other control variables are the number of organizations, endorsement policies, the Byzantine failures of organizations, and the number of organizations to which each organization gossips the transaction, which we refer to as the \textit{Gossip Ratio}. The gossips are propagated at one-second intervals. For the endorsement policies of $\mathit{EP: \{ q \; of \; n \} }$, the clients send the proposals and transactions to exactly $\mathit{q}$ organizations. On Fabric and FabricCRDT, organizations can contain several nodes, also known as \textit{peers}. In our experiments, each organization of these systems consists of one peer. Additionally, the blocks created by the Fabric's and FabricCRDT's ordering service have a size of 50 transactions, as based on our investigation and other studies~\cite{jeeta, FabricCRDT}, this block size yields good performance. Each experiment is executed for 180 seconds.

For the synthetic application, we used 1000 clients. $\mathit{ObjCount}$, $\mathit{OpsPerObjCount}$, and $\mathit{CRDTType}$ are control variables. We defined 1000 voters, eight elections, and eight parties per election for the voting application. We defined 1000 bidders, eight auctions, and a gradually growing number of bids for the auction application. We chose these values according to the scalability evaluation of Fabric done by other authors~\cite{jeeta}. The input parameters for modify- and read-transactions are randomly selected from these predefined values based on a uniform distribution during the experiment.

Each experiment is executed at least three times, and the results are averaged. At the end of each experiment, the performance metrics are collected. We measure the \textit{transaction throughput}, the \textit{average transaction latency}, the \textit{1st percentile transaction latency} and the \textit{99th percentile transaction latency}. The transaction throughput is the total number of successfully committed transactions divided by the total time taken for committing these transactions. The transaction latency is the response time per transaction from sending the proposal until receiving the commit receipts from organizations, according to the endorsement policy.

\begin{figure*}[h!]
  \centering
	\input{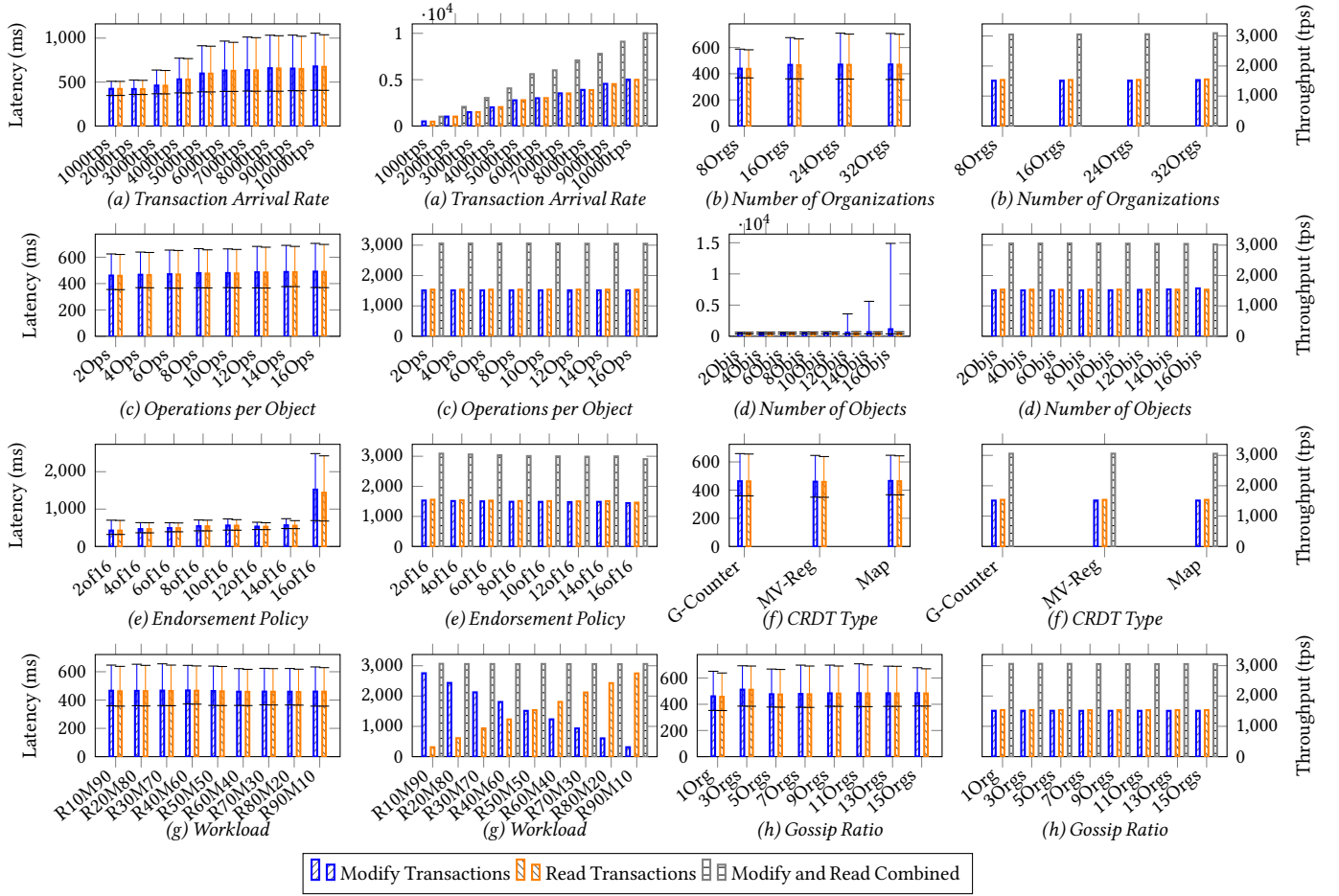}
  \caption{Throughput, average, 1st, and 99th percentiles transaction latencies for executed configurations of synthetic application on \orderlesschain.}
  \label{fig:sync}
\end{figure*}

\textbf{Experimental Setup --} Each organization of \orderlesschain, Fabric, and FabricCRDT runs on an individual KVM-based Ubuntu 20.04 virtual machine (VM), and different organizations do not share VM resources. Each VM uses 9.8 GB of RAM and four vCPUs. Since the VMs are located within a single cluster and are connected via LAN,  we used Ubuntu's \textit{NetEm (network emulation)} and \textit{tc (traffic control)} facilities for adding 100 ms delay, 4 ms jitter, and 100 Mbits rate control to all links for emulating a WAN. We chose these values by observing the delays and bandwidth between two Ubuntu servers in two different cities in Europe and North America, provided by two different cloud providers. The ordering service of Fabric and FabricCRDT runs on a separate VM. We also developed a distributed benchmarking tool that orchestrates a distributed deployment of clients, generates and submits transactions, and collects performance metrics.

\begin{table}[h!]
	\scriptsize
	\centering
	\caption{Control variables of synthetic application.}
	\begin{tabular}{| l l l |} 
		\hline
		\textbf{Control Variable} & \textbf{Default Value}           & \textbf{Executed Configuration}                                            \\ 
		\hline
		\hline
		(1) TS Arrival Rate       & 3000 tps                         & \{1000 tps, ..., 10,000 tps\}                                              \\ 
		(2) Number of Orgs        & 16  Orgs                         & \{8 Orgs, ..., 32 Orgs\}                                                   \\ 
		(3) Operations per Obj           & 1 Op                             & \{2 Ops, ..., 16 Ops\}                                                     \\
		(4) Number of Obj         & 1 Obj                            & \{2 Objs, ..., 16 Objs\}                                                   \\
		(5) Endorsement Policy    & $\mathit{ \{ 4 \; of \; 16 \} }$ & \{$\mathit{\{ 2 \; of \; 16 \} }$, ..., $\mathit{\{ 16 \; of \; 16 \} }$\} \\ 
		(6) CRDT Type             & G-Counter                        & \{G-Counter, MV-Register, Map\}                                            \\
		(7) Workload (Read/Mdfy)   & R50M50                           & \{R10M90, ..., R90M10\}                                                    \\ 
		(8) Gossip Ratio          & 1 Org                            & \{1 Org, ..., 15 Orgs\}                                                    \\ 
		(9) Byzantine Orgs       & 0 Failure                        & \{1 Failure, 2 Failures, 3 Failures\}                                      \\ 
		\hline
	\end{tabular}
	\label{table:control_vars_sync}
\end{table}

\textbf{Experimental Results for Synthetic Application on \orderlesschainspace --} Table~\ref{table:control_vars_sync} displays the control variables, their default values, and the executed experimental configurations for the synthetic application on \orderlesschain. \begin{wrapfigure}{l}{4.5cm}
  \begin{center}
    \pgfplotstableread[row sep=\\,col sep=&]{
	throughput & latency & latencytop & latencybottom \\
	 1000 &  412 &  100 &  67    \\
	 2000 &  418 &  100 &  60    \\
	 3000 &  439 &  119 &  62    \\
	 4000 &  508 &  194 &  118    \\
	 5000 &  589 &  208 &  192    \\
	 6000 &  600 &  232 &  185    \\
	 7000 &  646 &  273 &  216    \\
	 8000 &  713 &  305 &  269    \\
	 9000 &  1105 &  775 &  547    \\
	 10000 &  7984 &  13378 &  7320    \\
	  }\scalezerothroughputlatencyall

\pgfplotstableread[row sep=\\,col sep=&]{
	throughput & latency & latencytop & latencybottom \\
	 1000 &  422 &  86 &  72    \\
	 2000 &  420 &  100 &  64    \\
	 3000 &  459 &  164 &  96    \\
	 4000 &  538 &  224 &  165    \\
	 5000 &  618 &  321 &  226    \\
	 6000 &  626 &  344 &  234    \\
	 7000 &  642 &  358 &  239    \\
	 8000 &  652 &  381 &  256    \\
	 9000 &  656 &  357 &  252    \\
	 10000 &  680 &  364 &  269    \\
	  }\scaleonethroughputlatencyall

\pgfplotstableread[row sep=\\,col sep=&]{
	throughput & latency & latencytop & latencybottom \\
	 1000 &  435 &  74 &  89    \\
	 2000 &  425 &  111 &  75    \\
	 3000 &  471 &  264 &  115    \\
	 4000 &  525 &  286 &  163    \\
	 5000 &  576 &  339 &  198    \\
	 6000 &  598 &  347 &  217    \\
	 7000 &  618 &  372 &  231    \\
	 8000 &  624 &  381 &  235    \\
	 9000 &  650 &  399 &  263    \\
	 10000 &  654 &  396 &  262    \\
	  }\scaletwothroughputlatencyall

\pgfplotstableread[row sep=\\,col sep=&]{
	throughput & latency & latencytop & latencybottom \\
	 1000 &  447 &  70 &  91    \\
	 2000 &  437 &  180 &  90    \\
	 3000 &  468 &  273 &  115    \\
	 4000 &  510 &  311 &  147    \\
	 5000 &  564 &  343 &  193    \\
	 6000 &  560 &  354 &  198    \\
	 7000 &  594 &  374 &  216    \\
	 8000 &  593 &  377 &  219    \\
	 9000 &  613 &  388 &  236    \\
	 10000 &  625 &  392 &  238    \\
	  }\scalethreethroughputlatencyall

\begin{tikzpicture}[ font=\footnotesize, ]
	\begin{axis}[
			ylabel=Latency (ms),
			xlabel=Throughput (tps),
			xmin=0,
			ymin=0,
			height=4cm,
			ymajorgrids=true,
			grid style={blue!15},
			clip=false,
			legend pos=south east
		]

        
		\addplot plot [color=olive, mark=x] table[x=throughput,y=latency]{\scaleonethroughputlatencyall};
		\addlegendentry{ 16 Orgs };
        
		\addplot plot [color=blue, mark=+] table[x=throughput,y=latency]{\scaletwothroughputlatencyall};
		\addlegendentry{ 24 Orgs };
        
		\addplot plot [color=brown, mark=|] table[x=throughput,y=latency]{\scalethreethroughputlatencyall};
		\addlegendentry{ 32 Orgs };

	\end{axis}

\end{tikzpicture}
  \end{center}
   \caption{Average latency to throughput for an increasing number of organizations.}
  \label{fig:scale}
\end{wrapfigure}
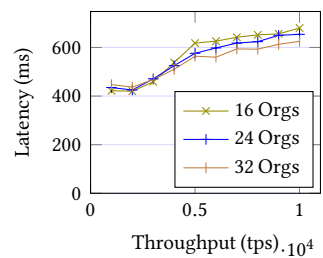One of the control variables is set to the executed configurations for each experiment, and the other control variables are set to the default value. The results of the experiments are shown in Figure~\ref{fig:sync}. As shown in Figure~\ref{fig:sync}(a), we observe that the throughput increases with an increasing transaction arrival rate. However, the latency increases as the load on the system increases. We studied the effect of increasing the number of organizations on throughput and latency, as shown in Figure~\ref{fig:sync}(b). For each experiment, we set the endorsement policy to $\mathit{EP: \{ 4 \; of \; NumerOfOrgs \} }$. We observe that the system scales for increasing organizations without affecting the throughput and latency. Furthermore, as shown in Figure~\ref{fig:scale}, we compared the average latency to throughput for an increasing number of organizations and arrival rates and observed that \orderlesschainspace scales. Figure~\ref{fig:sync}(c) shows that throughput and latency are unaffected by the increasing number of operations. However, in Figure~\ref{fig:sync}(d), the latency increases for a larger number of objects in the transaction. The reason is that we lock the objects in the cache to avoid concurrent reads and writes while applying the modify-transactions. With an increasing number of organizations required by the endorsement policy, we observe that the latency increases as the load on the organization increases, as  shown in Figure~\ref{fig:sync}(e). We observed that latency and throughput are independent of CRDT types, demonstrated in Figure~\ref{fig:sync}(f). In Figure~\ref{fig:sync}(g), we gradually decreased the modify-transactions in the workload from 90 percent to 10 percent, and we observed that the latency and throughput were not affected.  As demonstrated in Figure~\ref{fig:sync}(h), we did not observe a significant change in latency and throughput for an increasing gossip ratio either.

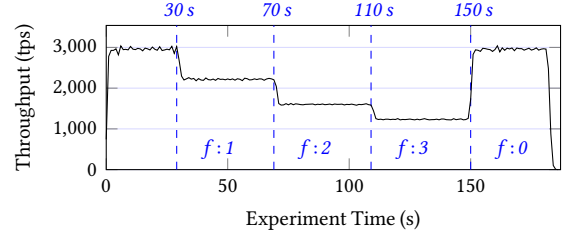
\begin{figure}
  \centering
	\begin{tikzpicture}[ font=\footnotesize, ]
	\begin{axis}[
			ylabel=Throughput (tps),
			xlabel=Experiment Time (s),
			xmin=0,
			ymin=0,
			ymax=3539,
			xmax=187,
		    width=.9\linewidth,
			height=3.5cm,
			ymajorgrids=true,
			grid style={blue!15},
			clip=false,
		]


        \draw [dashed, color=blue] (29,0) -- (29,3539);
        \node[above,blue] at (46,20) {$\mathit{f : 1 }$};
        \node[above,blue] at (29,3539) {\textit{ 30 s}};

        \draw [dashed, color=blue] (69,0) -- (69,3539);
        \node[above,blue] at (86,20) {$\mathit{f : 2 }$};
        \node[above,blue] at (69,3539) {\textit{ 70 s}};

        \draw [dashed, color=blue] (109,0) -- (109,3539);
        \node[above,blue] at (126,20) {$\mathit{f : 3 }$};
        \node[above,blue] at (109,3539) {\textit{ 110 s}};

        \draw [dashed, color=blue] (150,0) -- (150,3539);
        \node[above,blue] at (167,20) {$\mathit{f : 0}$};
        \node[above,blue] at (150,3539) {\textit{ 150 s}};

		\addplot[color=black,] coordinates {
			 (0, 742)  (1, 2779)  (2, 2926)  (3, 2927)  (4, 2966)  (5, 2818)  (6, 3037)  (7, 2945)  (8, 2952)  (9, 2939)  (10, 3023)  (11, 2956)  (12, 2946)  (13, 2976)  (14, 2963)  (15, 2916)  (16, 2983)  (17, 2965)  (18, 2937)  (19, 3008)  (20, 2954)  (21, 2930)  (22, 3017)  (23, 2923)  (24, 2937)  (25, 2956)  (26, 2941)  (27, 3026)  (28, 2853)  (29, 3019)  (30, 2682)  (31, 2295)  (32, 2200)  (33, 2228)  (34, 2258)  (35, 2221)  (36, 2238)  (37, 2181)  (38, 2234)  (39, 2200)  (40, 2191)  (41, 2250)  (42, 2209)  (43, 2201)  (44, 2223)  (45, 2192)  (46, 2237)  (47, 2215)  (48, 2220)  (49, 2204)  (50, 2226)  (51, 2207)  (52, 2227)  (53, 2196)  (54, 2230)  (55, 2218)  (56, 2185)  (57, 2244)  (58, 2232)  (59, 2201)  (60, 2207)  (61, 2228)  (62, 2211)  (63, 2228)  (64, 2225)  (65, 2193)  (66, 2225)  (67, 2240)  (68, 2228)  (69, 2205)  (70, 2043)  (71, 1622)  (72, 1584)  (73, 1589)  (74, 1619)  (75, 1591)  (76, 1623)  (77, 1593)  (78, 1615)  (79, 1607)  (80, 1596)  (81, 1591)  (82, 1602)  (83, 1604)  (84, 1607)  (85, 1611)  (86, 1591)  (87, 1605)  (88, 1598)  (89, 1611)  (90, 1594)  (91, 1600)  (92, 1609)  (93, 1592)  (94, 1610)  (95, 1602)  (96, 1603)  (97, 1608)  (98, 1588)  (99, 1602)  (100, 1597)  (101, 1608)  (102, 1595)  (103, 1581)  (104, 1601)  (105, 1595)  (106, 1604)  (107, 1617)  (108, 1594)  (109, 1585)  (110, 1444)  (111, 1251)  (112, 1231)  (113, 1238)  (114, 1224)  (115, 1238)  (116, 1238)  (117, 1228)  (118, 1236)  (119, 1230)  (120, 1252)  (121, 1221)  (122, 1232)  (123, 1228)  (124, 1234)  (125, 1241)  (126, 1230)  (127, 1222)  (128, 1238)  (129, 1223)  (130, 1237)  (131, 1251)  (132, 1233)  (133, 1222)  (134, 1227)  (135, 1228)  (136, 1238)  (137, 1223)  (138, 1243)  (139, 1242)  (140, 1226)  (141, 1223)  (142, 1220)  (143, 1250)  (144, 1226)  (145, 1236)  (146, 1246)  (147, 1222)  (148, 1229)  (149, 1259)  (150, 1726)  (151, 2827)  (152, 2940)  (153, 2930)  (154, 2902)  (155, 2978)  (156, 2970)  (157, 2955)  (158, 2945)  (159, 2901)  (160, 2947)  (161, 3039)  (162, 2890)  (163, 2989)  (164, 2928)  (165, 2981)  (166, 2962)  (167, 2986)  (168, 2919)  (169, 2984)  (170, 2969)  (171, 2970)  (172, 2932)  (173, 2957)  (174, 2980)  (175, 2934)  (176, 2953)  (177, 2989)  (178, 2933)  (179, 2957)  (180, 2967)  (181, 2959)  (182, 2480)  (183, 878)  (184, 102)  (185, 6)  (186, 1) 
		};

	\end{axis}

\end{tikzpicture}
  \caption{Throughput across experiments with Byzantine organizations.}
  \label{fig:byzantine_1}
\end{figure}
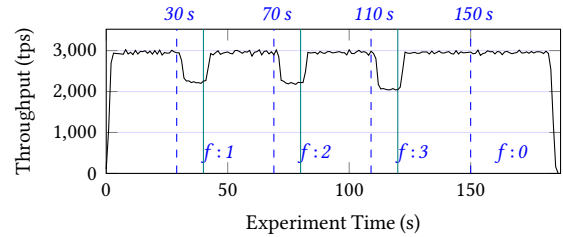
\begin{figure}
  \centering
	\begin{tikzpicture}[ font=\footnotesize, ]
	\begin{axis}[
			ylabel=Throughput (tps),
			xlabel=Experiment Time (s),
			xmin=0,
			ymin=0,
			ymax=3522,
			xmax=187,
		    width=.9\linewidth,
			height=3.5cm,
			ymajorgrids=true,
			grid style={blue!15},
			clip=false,
		]


        \draw [dashed, color=blue] (29,0) -- (29,3522);
        \node[above,blue] at (46,20) {$\mathit{f : 1 }$};
        \node[above,blue] at (29,3522) {\textit{ 30 s}};

        \draw [dashed, color=blue] (69,0) -- (69,3522);
        \node[above,blue] at (86,20) {$\mathit{f : 2 }$};
        \node[above,blue] at (69,3522) {\textit{ 70 s}};

        \draw [dashed, color=blue] (109,0) -- (109,3522);
        \node[above,blue] at (126,20) {$\mathit{f : 3 }$};
        \node[above,blue] at (109,3522) {\textit{ 110 s}};

        \draw [dashed, color=blue] (150,0) -- (150,3522);
        \node[above,blue] at (167,20) {$\mathit{f : 0}$};
        \node[above,blue] at (150,3522) {\textit{ 150 s}};

        \draw [solid, color=teal] (40,0) -- (40,3522);
        \draw [solid, color=teal] (80,0) -- (80,3522);
        \draw [solid, color=teal] (120,0) -- (120,3522);

		\addplot[color=black,] coordinates {
			 (0, 108)  (1, 1086)  (2, 2698)  (3, 2938)  (4, 2929)  (5, 2926)  (6, 2919)  (7, 2910)  (8, 3022)  (9, 2918)  (10, 2958)  (11, 2903)  (12, 3005)  (13, 2918)  (14, 2978)  (15, 2975)  (16, 2995)  (17, 2934)  (18, 2986)  (19, 2913)  (20, 2999)  (21, 2915)  (22, 3012)  (23, 2892)  (24, 2953)  (25, 2938)  (26, 2940)  (27, 3007)  (28, 2969)  (29, 2946)  (30, 2947)  (31, 2760)  (32, 2328)  (33, 2257)  (34, 2261)  (35, 2219)  (36, 2240)  (37, 2214)  (38, 2217)  (39, 2200)  (40, 2254)  (41, 2268)  (42, 2591)  (43, 2927)  (44, 2895)  (45, 2949)  (46, 2995)  (47, 2908)  (48, 2937)  (49, 2954)  (50, 2966)  (51, 2932)  (52, 2921)  (53, 3002)  (54, 3003)  (55, 2972)  (56, 2934)  (57, 2975)  (58, 2962)  (59, 3005)  (60, 2865)  (61, 2959)  (62, 2972)  (63, 3018)  (64, 2895)  (65, 2979)  (66, 2993)  (67, 2920)  (68, 2971)  (69, 2898)  (70, 3015)  (71, 2855)  (72, 2275)  (73, 2252)  (74, 2206)  (75, 2187)  (76, 2221)  (77, 2198)  (78, 2175)  (79, 2226)  (80, 2217)  (81, 2220)  (82, 2480)  (83, 2956)  (84, 2942)  (85, 2948)  (86, 2957)  (87, 2915)  (88, 2979)  (89, 2943)  (90, 2931)  (91, 2973)  (92, 2977)  (93, 2945)  (94, 2950)  (95, 2964)  (96, 2974)  (97, 2947)  (98, 2951)  (99, 2894)  (100, 3019)  (101, 2930)  (102, 2926)  (103, 2970)  (104, 3007)  (105, 2957)  (106, 2974)  (107, 2954)  (108, 2886)  (109, 2984)  (110, 2932)  (111, 2803)  (112, 2172)  (113, 2064)  (114, 2073)  (115, 2055)  (116, 2041)  (117, 2044)  (118, 2073)  (119, 2041)  (120, 2044)  (121, 2106)  (122, 2517)  (123, 2963)  (124, 2939)  (125, 2932)  (126, 2947)  (127, 2956)  (128, 2933)  (129, 2935)  (130, 2966)  (131, 2966)  (132, 2966)  (133, 2916)  (134, 2987)  (135, 2926)  (136, 2953)  (137, 2989)  (138, 2944)  (139, 2940)  (140, 2960)  (141, 2975)  (142, 2927)  (143, 2916)  (144, 2998)  (145, 2918)  (146, 2970)  (147, 2945)  (148, 2951)  (149, 2957)  (150, 2964)  (151, 2922)  (152, 2934)  (153, 3012)  (154, 2883)  (155, 2925)  (156, 2944)  (157, 2996)  (158, 2971)  (159, 2977)  (160, 2920)  (161, 2978)  (162, 2967)  (163, 2900)  (164, 2968)  (165, 2980)  (166, 2947)  (167, 2924)  (168, 2942)  (169, 2952)  (170, 3011)  (171, 2943)  (172, 2943)  (173, 2969)  (174, 2935)  (175, 2978)  (176, 2942)  (177, 2969)  (178, 2972)  (179, 2963)  (180, 2929)  (181, 2985)  (182, 2932)  (183, 2542)  (184, 1243)  (185, 160)  (186, 2) 
		};

	\end{axis}

\end{tikzpicture}
  \caption{Throughput across experiments with clients avoiding Byzantine organizations.}
  \label{fig:byzantine_2}
\end{figure}

Finally, we studied the effects of organizations' Byzantine failures. As shown in Figure~\ref{fig:byzantine_1}, three randomly selected organizations behaved arbitrarily for a specific period during the experiment. The Byzantine organizations either randomly avoid responding to clients' proposals or transactions or endorse the proposals incorrectly. The Byzantine organizations also randomly avoid forwarding the transactions to other organizations. We included three Byzantine organizations as, based on the $\mathit{EP: \{ 4 \; of \; 16 \} }$, the safety and liveness of the application can tolerate up to three Byzantine failures. We observed that the throughput decreases with every Byzantine failure. However, the latency is not affected. The reason for the decreasing throughput is that clients cannot collect an adequate number of endorsements due to the Byzantine organization not responding and the signature validation failure caused by the wrongly endorsed proposals. Since clients can observe organizations that wrongly endorse or do not respond while other organizations respond with lower latency, they can avoid Byzantine organizations. To demonstrate this, we ran experiments where the clients avoided the Byzantine organization and changed to another randomly selected organization. As shown in Figure~\ref{fig:byzantine_2}, the throughput returns to its pre-failure value immediately after clients avoid the Byzantine organizations, as shown by the solid green lines. 

\begin{figure*}
  \centering
	\input{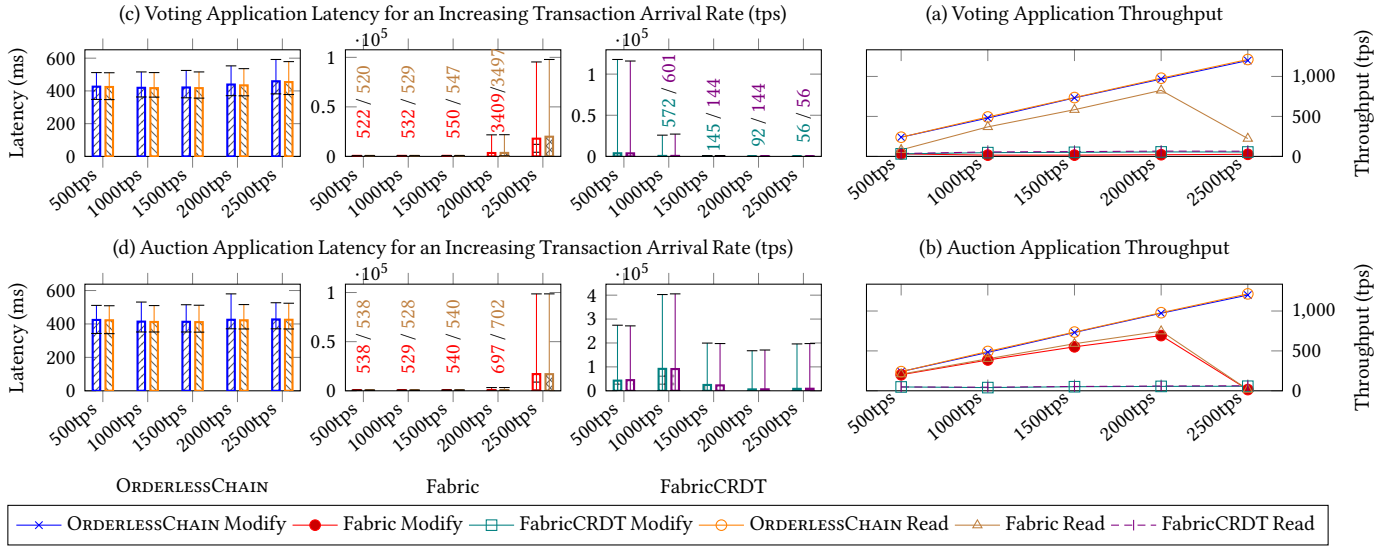}
  \caption{Experiments with voting and auction applications.}
  \label{fig:auction_vote}
\end{figure*}

\textbf{Vote and Auction Applications --} These experiments are conducted with eight organizations, the $\mathit{EP: \{ 4 \; of \; 8\} }$ endorsement policy, and the uniform workload. We increased the transaction arrival rate from 500 tps to 2500 tps. For FabricCRDT, we observed that latency significantly increases for a higher transaction arrival rate due to its CRDT implementation, so we limited the transaction latency for FabricCRDT to 240 seconds, after which they are timed out and not considered for throughput and latency determination. The results of the experiments are shown in Figure~\ref{fig:auction_vote} for \orderlesschain, Fabric, and FabricCRDT. As shown in Figure~\ref{fig:auction_vote}(a) for the voting application and Figure~\ref{fig:auction_vote}(b) for the auction application, we observe that \orderlesschainspace demonstrates a higher throughput of modify- and read-transactions for both applications. On Fabric, the failed transactions due to the MVCC validation explain its low throughput. Although we used caching for the CRDT approach in FabricCRDT, their approach still is a bottleneck. As shown in Figure~\ref{fig:auction_vote}(c) and Figure~\ref{fig:auction_vote}(d) (for the lower values, the average latencies are written on the plots), \orderlesschain's latency remains constant under increasing arrival rates. Fabric's latency significantly increases for higher arrival rates. The reason is that Fabric's central ordering service is a bottleneck. The increased latency causes more transactions to fail due to MVCC validation, which explains the significant throughput decrease for Fabric, especially from a 2000 tps to a 2500 tps arrival rate, as shown in Figure~\ref{fig:auction_vote}(b). FabricCRDT demonstrates irregular latency patterns as timed-out transactions are not considered. 

\textbf{Discussion --} We initiated this work to study whether permissioned blockchains need total global order of transactions. As demonstrated, the answer depends on the applications running on the permissioned blockchains. Suppose the invariant conditions can be modeled as I-confluent invariants. In that case, coordination is unnecessary, and our approach can be used to improve throughput and latency compared to coordination-based approaches significantly. However, coordination to order the transactions is required for applications with non-I-confluent invariants. For example, specifying a deadline for the end of an election or an auction and rejecting the transactions that arrive after this deadline is a non-I-confluent invariant and requires coordination to be preserved. One approach for enabling \orderlesschainspace to preserve such invariants is extending our system with coordination-based protocols such as the protocol used by Fabric and using this protocol when required. For example, the coordination-based protocol can be enabled when we are close to the end of an election or auction; otherwise, we use our coordination-free approach. There exist several I-confluent CRDT-based use cases~\cite{uc_1, crdt_problem_2, uc_3, uc_4, uc_5, uc_6, uc_7, orderlesschain_demo, orderlessfl, orderlessfile}, from key-value stores to multi-user collaborative environments, which can be implemented on \orderlesschain, benefiting from the trust and scalability our system offers. We implemented additional use cases on \orderlesschainspace (not evaluated in this paper) as proof of concept. We implemented an IoT-based supply chain use case to monitor the health of temperature-sensitive products during transit. We also implemented a trusted distributed file storage and a private and distributed machine learning environment by extending \orderlesschainspace with customized CRDTs~\cite{orderlesschain_demo, orderlessfl, orderlessfile}. The development of these applications on \orderlesschainspace was relatively straightforward. 


\section{Related Work} \label{rw_work}

The high computational overhead and low throughput of Proof-of-Work-based protocols make them infeasible for permissioned blockchains~\cite{pow_problems}. Hence, permissioned blockchains such as R3 Corda~\cite{corda} and Fabric~\cite{mainfabric} make use of coordination-based consensus protocols. R3 Corda uses a combination of single notaries, Raft~\cite{raft} and BFT-SMaRt~\cite{bftsmart}. Fabric currently uses a Raft-based protocol. Although these protocols improve performance, the required coordination among blockchain nodes negatively affects performance. Furthermore, the Raft-based ordering service of Fabric is not BFT. Other studies proposed BFT ordering services for Fabric~\cite{bft_fabric_2, bft_fabric_1}. 

Li et al.~\cite{Fast_as_Possible} proposed preserving application-level consistency by offering global coordination only when the application requires strong consistency. Otherwise, their proposed solution uses a faster, eventually consistent method with less coordination for a weaker consistency. However, this is designed for a non-Byzantine environment. Kleppmann and Howard~\cite{martin_bft} introduced an approach for processing I-confluent transactions. The authors introduced a BFT eventually consistent replicated database and proposed an approach for creating a directed acyclic graph (DAG)-based dependency graph of transactions. Non-faulty nodes periodically retrieve the missing dependent transactions from other non-faulty organizations. However, their work focuses on peer-to-peer databases and does not offer an environment for the trusted execution of decentralized applications.   

CRDTs have shown to be a valuable tool for managing concurrent and conflicting updates in several distributed systems~\cite{Redis, Riak, crdtdb, uc_13, Yunhao}. Some studies proposed coordination-based BFT approaches for executing CRDT applications~\cite{bft_crdt_1, bft_crdt_2, bft_crdt_3, bft_crdt_4, bft_crdt_5}. However, there have only been limited works studying CRDTs in blockchains. Vegvisir~\cite{dag_crdt} studied integrating CRDTs with a DAG-structured blockchain; however, it does not support executing smart contracts. FabricCRDT~\cite{FabricCRDT} is a permissioned blockchain using JSON CRDT techniques. As an extension of Fabric, FabricCRDT also uses a coordination-based protocol. The main difference between the CRDT approach used in FabricCRDT and \orderlesschainspace is that for every modification on FabricCRDT, the entire object stored on the ledger must be retrieved and modified in the smart contract and then be sent to organizations to be merged with the existing objects on the ledger. On FabricCRDT, the objects gradually become prohibitively large, negatively affecting the performance, as observed in here. 

\section{Conclusions} \label{conclusion}

We presented \orderlesschain, a coordination-free permissioned blockchain capable of running safe and live CRDT-based I-confluent applications in a Byzantine environment. Our evaluation shows that a coordination-free permissioned blockchain performs better than coordination-based ones.

\bibliographystyle{ACM-Reference-Format}
\bibliography{bibliography}

\end{document}